\documentstyle[12pt]{article}
\begin{document}
\topmargin 0cm
\baselineskip=.7cm
\parskip=.2cm
\textwidth=14cm
\textheight=21cm
\def\theequation{\arabic{section}.\arabic{equation}}
\newcommand{\beq}{\begin{equation}}
\newcommand{\eeq}{\end{equation}}
\newcommand{\rbracket}{\right]}
\newcommand{\fff}{{\bar f}}
\newcommand{\lbracket}{\left[}
\newcommand{\DD}{{\cal D}}
\newcommand{\DDt}{{\nabla}_{t}}
\newcommand{\ddt}{\partial_t}
\newcommand{\ddx}{\partial_x}
\newcommand{\h}{\theta}
\newcommand{\r}{\cal G}
\newcommand{\curve}{\Sigma_{\tau}}
\newcommand{\intc}{\int_{\Sigma_{\tau}}}
\newcommand{\Ag}{A^{g}}
\newcommand{\aaa}{{\alpha}}
\newcommand{\Q}{\bar{Q}}
\newcommand{\haha}{\theta_{11}}
\newcommand{\Pb}{\bar{P}}
\newcommand{\ppb}{\bar p_{i}}
\newcommand{\ps}{\bar{\psi}}
\newcommand{\LL}{{\cal L}}
\newcommand{\ggg}{\sf g}
\newcommand{\hgg}{\hat g}
\newcommand{\elg}{{\hat {\sf g}}^{\Sigma_{\tau}}}
\newcommand{\eld}{({\hat {\sf g}}^{\Sigma_{\tau}})^{*}}
\newcommand{\Rn}{{\cal R}_{\nu}}
\newcommand{\ppz}{\phi (z, {\bar z})}
\newcommand{\aaz}{{\bar A} (z, {\bar z})}
\newcommand{\ggz}{g (z, {\bar z}) }
\newcommand{\slnc}{{\it sl}_{N} ({\bf C})}
\newcommand{\SLnc}{{\it SL}_{N} ({\bf C})}
\newcommand{\sun}{{\it su}_{N}}
\newcommand{\ep}{\epsilon_{i}}
\newcommand{\Vq}{\Delta (\Q)}
\newcommand{\Vp}{\Delta (\Pb)}
\newcommand{\rw}{\rightarrow}
\newcommand{\lll}{\lambda}
\def\op{operator}
\def\tly{topologically}
\def\mfti{Moscow Institute of Physics and Technology}
\def\onlabs{permanent address: $\;$ }
\def\itep{Institute of Theoretical and Experimental Physics}
\def\Clgr{Calogero $\;$}
\def\CG{Clebsh-Gordon $\;$}
\def\gl{global}
\def\an{anomaly}
\def\f{field}
\def\exst{existence}
\def\nl{normalizable}
\def\nty{normalizability}
\def\cn{condition}
\def\con{configuration}
\title {Elliptic Calogero-Moser System From Two Dimensional Current Algebra}
\author { \sf Alexander Gorsky and Nikita Nekrasov \\
\centerline {\em  \itep}\\
\centerline{\em 117259,B.Cheremushkinskaya,Moscow,Russia}\\
\thanks {email: gorsky@vxitep.itep.msk.su    nikita@vxitep.itep.msk.su}
\centerline{\em ITEP-NG1/94    ,      hepth/9401ddd}
}
\maketitle
\begin{abstract}
We show that elliptic Calogero-Moser system and its Lax operator
found by Krichever can be obtained by hamiltonian reduction from
the integrable hamiltonian system on the cotangent bundle to the central
extension of the algebra of $\slnc$ currents.

\end{abstract}

\newpage
\section{Two Dimensional Current Algebra}
In this section we review relevant for us properties of two dimensional
current algebra (all this can be found in \cite{ef} and in references
therein).
We consider the Lie algebra $\mbox{Maps}( {\curve} , {\slnc} )$ of
$\slnc$ - valued
currents on
the elliptic curve $\curve$ with modular parameter $\tau$. This algebra
has a remarkable central extension, which is defined via the
$H^{(1,0)}({\curve})^{*}$-valued $2$-cocycle:
\beq
c (X, Y) =  \int_{\curve} \omega \wedge < X, dY >
\label{cocycle}
\eeq
$\omega \in H^{(1,0)}(\curve)$. Fix a holomorphic 1-differential $\omega$ with
the periods along the $A-$ and $B-$cycles on the $\curve$:
$$
\int_{A} \omega = 1, \; \int_{B} \omega = \tau
$$
Let us take a cotangent bundle $T^{*} \elg$ , which is a space of 4-tuples
$$
( \phi, c , {\bar A}, \kappa ), \; \phi :  \curve \to \slnc, \; {\bar A} \in
\Omega^{(0,1)}(\curve) \otimes \slnc, \; c, \kappa \in {\bf C}
$$
and the pairing between algebra $\elg$ and its dual $\eld$ in this
parameterization has a form
\beq < ({\bar A}, \kappa ); \; (\phi, c) > =
\kappa \cdot c + \intc \omega \; tr \phi {\bar A}
\label{pairing}
\eeq

On the cotangent bundle acts naturally a current group ${\it SL}_{N}({\bf
C})^{\curve}$:

\begin{eqnarray}
&& \ppz \to \ggz \ppz \ggz^{-1}
\nonumber\\
&& \aaz d{\bar z} \to  \ggz \aaz \ggz^{-1} + {\kappa}
\ggz {\bar \partial} \ggz^{-1}
\nonumber\\
&& \kappa \to \kappa, \; c \to c + \intc \omega \;tr (\phi g {\bar \partial} g
)
\label{graction}
\end{eqnarray}

This action preserves a symplectic form $\Omega$ on $T^{*}{\elg}$:

\beq
\Omega = \delta c \wedge \delta \kappa +
\intc \omega \; tr ( \delta \phi \wedge \delta {\bar A} )
\label{sympl}
\eeq

The moment map has the form
\beq
\mu = {\kappa} {\bar \partial}\phi + [{\bar A} , \phi ]
\label{moment}
\eeq

\section{Hamiltonian Reduction And Integrable Model}
For our purposes it is more convinient to consider slightly enlarged
symplectic manifold, namely
$$
T^{*}{\elg} \times {\cal O}_{\nu}^{-}
$$
where ${\cal O}_{\nu}^{-} = {\bf CP}^{N-1}$ with symplectic form
$\omega_{\nu} = - N{\nu} \Omega_{Fubini-Shtudi}$. We endow this symplectic
manifold with the structure of $\SLnc^{\curve}$ space. On the first factor
the group acts as before, and on the second one acts the finite dimensional
$\SLnc$ sitting at point $0 \in \curve$ in a usual way.  We denote the
homogeneus coordinates on ${\cal O}_{\nu}$ as $(f_{1} : \dots :
f_{N} )$
Now we wish to apply a hamiltonian reduction at zero level of moment map.
It amounts to
$$
\mu = {\sf i} {\nu} ( Id - f \otimes f^{+} ) \frac{\delta (z, {\bar
z}) dz \wedge d{\bar z}}{\omega}
$$

The generic element of $\elg$ by the action (\ref{graction}) can be
transformed
to semi-simple constant element of maximal torus of $\slnc$. Let us make this
transformation.  It is defined ambiguously, once $\bar A$ has a form
$\mbox{diag}(a_{j})$, it can be transformed (by means of large
gauge transformation) to
$\mbox{diag}(a_{j} + \omega_{m_{j} n_{j}})$,
$\omega_{m_{j} n_{j}} = m_{j} + \tau n_{j}$,
$(m_{j}, n_{j}) \in {\bf Z}^{2}$, so the real space of parameters,
which distinguish the coadjoint orbits, is the moduli space of flat
$\sun$-connections.
After applying this gauge transformation we arrive to equation, written
in terms of matrix elements:
$$
{\kappa} {\bar \partial}\phi_{ij} + a_{ij} \phi_{ij} = {\sf i}{\nu}
(\delta_{ij} - f_{i}f_{j}^{*} )
$$

here $a_{ij} = a_{i} - a_{j}$, $a_{i} \in
{\bf C}$ are diagonal entries of $\bar A$.

Diagonal component of this equation gives us two
kinds of constraints:
$$
1 - f_{i} f_{i}^{*} = 0
$$
and
$$
\phi_{ii} = p_{i} = const
$$
this comes from the fact that on the ellpitic curve
there are no meromorphic functions except constants
with the only one pole of the first order. These condition imply that
all ${\cal O}_{\nu}$ -like degrees of freedom are freezed (they become
dynamical in the situation with group of currents, like in \cite{gorrui}).

Non-diagonal components have the following form:

$$
\phi_{ij} = \exp ( \pi \frac{a_{ij} (z- {\bar
z})}{{\kappa} \tau_{2}} ) \psi_{ij}
$$

where $\psi_{ij}$ is a section of meromorphic line bundle over $\curve$
with one pole at the point $0$ of the first order and the following
monodromy properties:
$$
\psi_{ij} (z + 1) = \psi_{ij}(z)
$$
$$
\psi_{ij} (z + \tau) = e^{-\frac{2\pi {\sf i}}{\kappa} a_{ij}} \psi_{ij} (z)
$$
The solution has the following form:

\beq
\psi_{ij} (z) = \frac{\nu}{\kappa}  \frac{\haha (z + \frac{a_{ij}}{\kappa}
)}{\haha (z) \haha (\frac{a_{ij}}{\kappa} )}
\label{solution}
\eeq

It gives Krichever answer \cite{krichever}
for the Lax matrix. Invariants of the matrix $\ppz$ give us Hamiltonians for
integrable model. The first non-trivial one is
$$
\mbox{tr} \ppz^{2} = \sum_{i} \frac{1}{2} p_{i}^{2} +
\frac{{\nu}^{2}}{\kappa^{2}} \sum_{i < j} {\wp} (\frac{a_{ij}}{\kappa}) -
{\wp}(z)
$$

 It is a Hamitonian of elliptic Calogero-Moser model and as always due to
 the quantum corrections coupling constant ${\nu}^{2}$ gets shifted to
 $\nu(\nu -1)$.Note that elliptic Calogero-Moser system covers both
 periodical and nonperiodical Toda chains and lattices \cite{inoz} which can
 be considered as a particular limits of the system above.To perform
 reduction to Toda theory one should introduce new variables like
 $a_{i}=x_{i}+(j-1) \frac{b}{\kappa}$ rescale the coupling constant
 $\nu={\nu}_{0}e^{\frac{b}{\kappa}}$ and take the limit $b \to \infty$.

\section{Elliptic Deformation Of Two Dimensional Yang-Mills Theory}

We remind that trigonometric Calogero-Moser model (Sutherland model)
can be viewed as an effective theory for zero modes of the gauge field
on a cylinder with inserted Wilson line in
representation $R_{\nu}$ ($N\nu$-th symmetric power of the $N$-dimensional
fundamental representation of $SU(N)$, which
corresponds to the orbit ${\cal O}_{\nu}$ (\cite{gor2d},\cite{gorrui}).

In the same way we expect that some kind of elliptic
deformation of Yang-Mills theory should exist.
We propose the following action for this hypothetical theory:
\beq
S_{\tau} = \int_{\curve \times S} \omega \wedge \mbox{tr} (\phi
F_{t {\bar z}} - {\varepsilon} \phi^{2})
\label{ellaction}
\eeq
Here $F_{t {\bar z}} = \partial_{t} {\bar A} - {\bar \partial} A_{t} +
[A_{t}, {\bar A}]$ is a component of the gauge field strenght, $\omega = dz$ -
holomorphic
differential on $\curve$, $S$ is a time-like circle.
In the limit $Im \tau \to \infty$ only those modes of gauge fields survive
which give rise to the theory on a cylinder,
rather then on a three-dimensional manifold.Note also that if one adds
Chern-Simons action to the Lagrangian above the system of interacting
particles in the external magnetic field immediately appears according to
the standard analysys.

We hope to get some information about wave functions and
spectrum of this model via this field theory language.
Essentually what is needed here, is the proper analogue of the decomposition
of the regular representation of the group $\sf G$ as
${\sf G}_{L} \times {\sf G}_{R}$ - module:
$$
{\cal L}^{2} ({\sf G}) = \bigoplus \alpha \otimes \alpha^{*}
$$
for the loop group (these are the irreducible
representations of the loop group which propagate along the
torus, like in the situation with the affine Lie algebra
(trigonometric case) \cite{gorrui}, \cite{gor2d}
irreducible representations of the finite-dimensional
group propagated along the circle).
{}From the results of \cite{efquan} we understand that an appropriate
replacement of this decomposition looks like $$ \int_{{\bf C}^{N}} d{\lll}
M_{{\lll},\kappa} \otimes M_{{\lll},\kappa}^{*} $$ where $M_{{\lll},\kappa}$
is a Verma module with the highest weight $\lll$ and the level $\kappa$.
Finally, in the trigonometric case states were enumerated by
the invariants $Inv( \alpha \otimes \alpha^{*} \otimes R_{\nu})$
\cite{gorrui},\cite{gor2d},
what is the same as the space of intertwiners
$\Phi: \alpha \to \alpha \otimes R_{\nu}$. Path integral (\ref{ellaction})
suggest that the same statement holds in the ellpitic case, the only
difference  is that finite-dimensional representations $\alpha$ of
$SU(N)$ (or $\SLnc$) are replaced by the Verma modules.  We refer to
\cite{efquan} for elaborated treatment of this construction with
intertwiners.

Let us mention one another important observation.We would like recall the
elliptic Sklyanin algebras \cite{skly2} which appeared in description of
asymmetric spin systems with the nearest-neibouhrs interactions.This
quadratic algebra has a particular representation in which generators can be
realized as finite-difference operators with coefficients expressed in terms
of elliptic functions.These operators act in finite dimensional
representations with fixed spin j.If one considers the limiting proceedure
described in \cite{skly2} which results in differential operators with
elliptic coefficients and consider Casimir in this representation then two
particle elliptic Calogero-Moser Hamiltonian appears.Let us outline the
difference between interpretation of spectral parameter in spin systems and
Yang-Mills theory.In spin systems it is the coordinate on the parameter
space while in YM situation it is nothing but the coordinate on the world
sheet.The meaning of this difference as wee as the role of elliptic algebra
in deformed YM theory definitely can shed additional light on the structure
of the integrable systems in 3d.In particular the answer about the origin
of the generalization of Yang-Baxter equation for elliptic
Calogero-Moser system \cite{skly1} should be found.

To conclude,we derived elliptic Calogero-Mozer system starting from
elliptic deformation of Yang-Mills theory.We expect to get elliptic
Ruijsenaars' models \cite{ruij},which is the top system for the tower of
integrable Hamiltonians for interacting particles,as a reduction of the
integrable system on the cotangent bundle to current group in two dimensions
(see \cite{gorrui}).We also postpone many quations related with solitonic
interpretation,finite-gap solutions of KP hierarchy and possible matrix
model counterparts of our model.

\section{Acknowledegments}
We are deeply indepted to V.Fock,P.G.O.Freund, A.Losev, M.A.Olshanetsky,
A.Rosly, S.Shatashvili,A.Zabrodin and especially to V.Rubtsov for
stimulating discussions.

  Research of N.N. was partially supported  by Soros Foundation Grant awarded
by American Physical Society, French Mathematical Society Grant via fund
"ProMathematika" and Swedish Institute Grant.

  We thank for support also RFFI, grant \# 93-02-14365 .

\end{document}